\documentclass[12pt]{article}

\title{ Hard Diffraction and Unitarity}

\author{V.A. Petrov \\
{\small Institute for High Energy Physics,
Protvino, 142280 Russia}}

\date{}

\begin{document}

\maketitle
\begin{itemize}

\item Unitarity of the $S$-matrix which stems from the postulate of
asymptotic
completeness (see e.g.~[1]) refers to asymptotic states,
representing
physical particles. In quantum field-theoretic terms it means that
one
deals with
on-shell, truncated Green functions.
Unitarity is tightly related to (but not exhausted by)
probabilistic
interpretation of the scattering and production amplitudes, and
effectively
prevents these amplitudes from too fast growth with energy~[2]. 

Hard processes in general, and hard diffraction in particular, are
often
related to off-shell amplitudes. Can unitarity, seemingly on-shell
property,
lead to limitations in this case also? In fact, unitarity of the
$S$-matrix, when
considered in the axiomatic framework, is assumed to hold off mass
shell~[3],
thereof, e.g., the optical theorem holds when ``external'' particles
are
virtual.

However the bounds which were proven for the on-shell case cannot be
derived
for more general off-shell case. 

This leads, in particular, to a possibility of a much faster rise
with energy
than in the on-shell case~[4]. 

\item In this talk we limit ourselves by consideration of deeply
inelastic
scattering (DIS)
at small~$x$, when it is believed to have mostly diffractive
character. 

In fact the most characteristic feature of a diffractive process is a
diffractive pattern of the scattered waves. In high-energy collisions
there is
a related feature, i.e. a rapidity gap between diffractively
scattered
(excited, produced) final states. Nonetheless only the study of
the diffractive
pattern can give us as information about global properties (size,
shape) of the
scatterer (``interaction region''). From intuitive considerations one
can think
that for off-shell scattering the interaction radius should decrease
with
growth of virtuality.

\item What is the role of unitarity? When asking such
a question
we mean the following. If one takes some ``bare'' or ``Born''
amplitude which is
deduced from some simple arguments (say, Regge pole)  it often
violates
unitarity or its consequences (e.g. Froissart--Martin (FM) bound).
This is
not the
reason to abandon such a ``wrong'' amplitude which is considered to
be very
good in many other respects.  The remedy is ``unitarization'', i.e.
some
infinite summation of the ``bare'' amplitude which yields a new, good
amplitude
respecting unitarity etc. The most known examples are eikonal and
U-matrix representations. 

Discovery of the fast growth of DIS cross-sections at HERA
exacerbated the quest of possible unitarity-driven upper bounds.

Such bounds were obtained (see e.g.[5]) but at a price, after making
serious extra assumptions which deprive the results of rigour and
generality  of the FM theorem.

\item However the framework of general principles of quantum
field theory seem to fairly admit power-like growth of the
"off-shell" cross-sections. Extended Regge-eikonal just realizes this
possibility in a concrete form. It is interesting to note that
cross-sections of exclusive (binary) deeply virtual processes do not
exceed the FM limit $(log^2s)$, while the corresponding  total
cross-sections grow as a power of energy [6]. It means that at
extremely  high $s$ and $Q^2$ "unitarity effects" for total
cross-sections are relatively negligible, while they are 100\%
important for binary exclusive cross-sections. 

A natural interpretation of this phenomenon is that at high $Q^2$ the
role of multiple  production grows in full accordance whis ref
[7].

We still lack the results concerning angular distribution of final
particles in the binary deeply virtual exclusive processes. At the
moment we can only mention the average impact parameter
$$
<b^2> (s,Q^2) \sim log^2 s / log Q^2.
$$

We see that the transverse interaction  region grows asymptotically 
with energy (feature familiar from on-shell hadron-hadron processes)
and shrinks with virtuality, $Q^2$, but slower.

At first sight it seems to mean that at equal c.m.s. energies
the off-shell diffractive pattern is shallower and has more wide
forward
peak. But at realistic $s$ and $Q^2$ the picture can be much  more
complicated. We have to stress that up to now no sign of a dip is
seen in the angular distribution of exclusively produced vector
mesons
at HERA. One could take this as an evidence in favour of a
$Q^2$-induced spread of the diffractive pattern.

\item  The last subject I want to touch  is the case when in
capacity of a hard scale we take not the virtuality but the
"compactification radius", $R_c"$, assuming in accordance with newest 
ideas that our space-time has more than 4 dimensions, and that extra
dimensions are somehow compactified. What  is the r\^ole of a hard
scale, $R_c$, in high energy behaviour? It appears that this  r\^ole
is quite insignificant.
At least for the upper bound. One can show [8], that the
compactification radius enters the upper bound (which is FM-like)
quite harmfully, and, with a proper normalization, peacefully 
disappears in the zero limit bringing us back to the usual Minkowsky
space-time and the FM bound. It is likely that 
influence of 
$R_c$ is more dramatic for
high momentum transfers.

\item As a conclusion I have no much to say.

1. Effects of fast growth of DIS cross-section discovered at HERA
remain unexplained.

2. Unitarity does not limit this growth too stringently.

I express my deep gratitude to organizers of the magnificent
workshop "Diffraction 2000" in Cetraro, especially to Roberto Fiore,
Alessandro Papa, and Enrico Predazzi, for their kind hospitality  and
valuable support.
\end{itemize}

\end{document}